# Earthquakes and cluster dynamics during Interseismic phases between the Northern and Central Apennines (Italy)

**Authors:** Marion Baques[1], Piero Poli[1], Michele Fondriest[1]

**Affiliations:**

1. Dipartimento di Geoscienze, Università degli Studi di Padova, Padova, Italy.

**Corresponding author:** marion.baques@unipd.it

**Abstract**

In the last thirty years, the Northern and Central Apennines (Italy) have been affected by three main destructive seismic sequences: the 1997 Colfiorito (three events ML> 5.5), the 2009 L'Aquila (one event ML>5.5), and the 2016-2017 Amatrice-Visso-Norcia (three events ML>5.5). Several studies have analysed the spatial-temporal evolution and processes driving each sequence, focused more on the foreshock-mainshock-aftershocks periods. Here, we focus on the 2018-2024 Interseismic phase, aiming to unravel the long-term seismogenic behaviour of this region. We first relocated the earthquake catalogue and identified clusters through a de-clustering algorithm. During this phase, background seismicity and most of the clusters were arranged in a 2-3 km thick low- angle layer. We found that (i) most clusters were driven by an aseismic process, (ii) the depth of both clusters and seismicity layer increased toward the southeast, (iii) the volume of clusters decreases to the southeast, (iv) the low-angle layer almost disappeared in the L'Aquila area. Comparing two Interseismic phases (2011-2016 and 2018-2024), we found striking similarities with events occurring on same sites and characterized by both foreshock-mainshock-aftershocks and swarm-like behaviour. In addition to this, the L'Aquila area was seismically more "silent" compared to the norther sites during both Interseismic phases. We propose that these different long-term seismogenetic behaviours may reflect variation in the structure and rheology of the upper crust moving from the Northern to the Central Apennines. This highlights the important role of the structural inheritance in controlling how the active deformation affects the Interseismic period.

## 1. Introduction

The Northern and Central Apennines (NCA) are one of the most seismically active areas in Italy and in all Europe, frequently experiencing destructive seismic sequences with mainshocks of magnitudes larger than 5. Notable main recent events include the Mw 5.7 Norcia earthquake in 1979 (Deschamps, Iannaccone & Scarpa, 1984), Mw 6.0 Colfiorito in 1997 (Deschamps et al., 2000), Mw 6.3 L'Aquila in 2009 (Chiarabba et al., 2009), and Mw 6.0, 5.9, and 6.5 Amatrice-Visso-Norcia (AVN) in 2016 (Marzocchi, Taroni & Falcone, 2017) (Fig. 1). The Colfiorito sequence started the 26th of September 1997 and stopped on the 3rd of November 1997 (light blue ellipse, Fig. 2a). It produced more than 2000 events, with 12 events with a magnitude greater than 4 (Deschamps et al., 2000). The L'Aquila sequence (blue ellipse, Fig. 2a), beginning in January 2009, generated over 5000 events before the Mw 6.1 mainshock on April 6th, 2009, followed by more than 50,000 aftershocks (Valoroso et al., 2013). Seven years later, the AVN sequence occurred (purple ellipse, Fig. 2a), starting on August 24th 2016 with an Mw 6 event, culminated in an Mw 6.5 earthquake on October 30th, 2016, with over 450,000 events recorded in the year following the mainshock (Spallarossa et al., 2021).

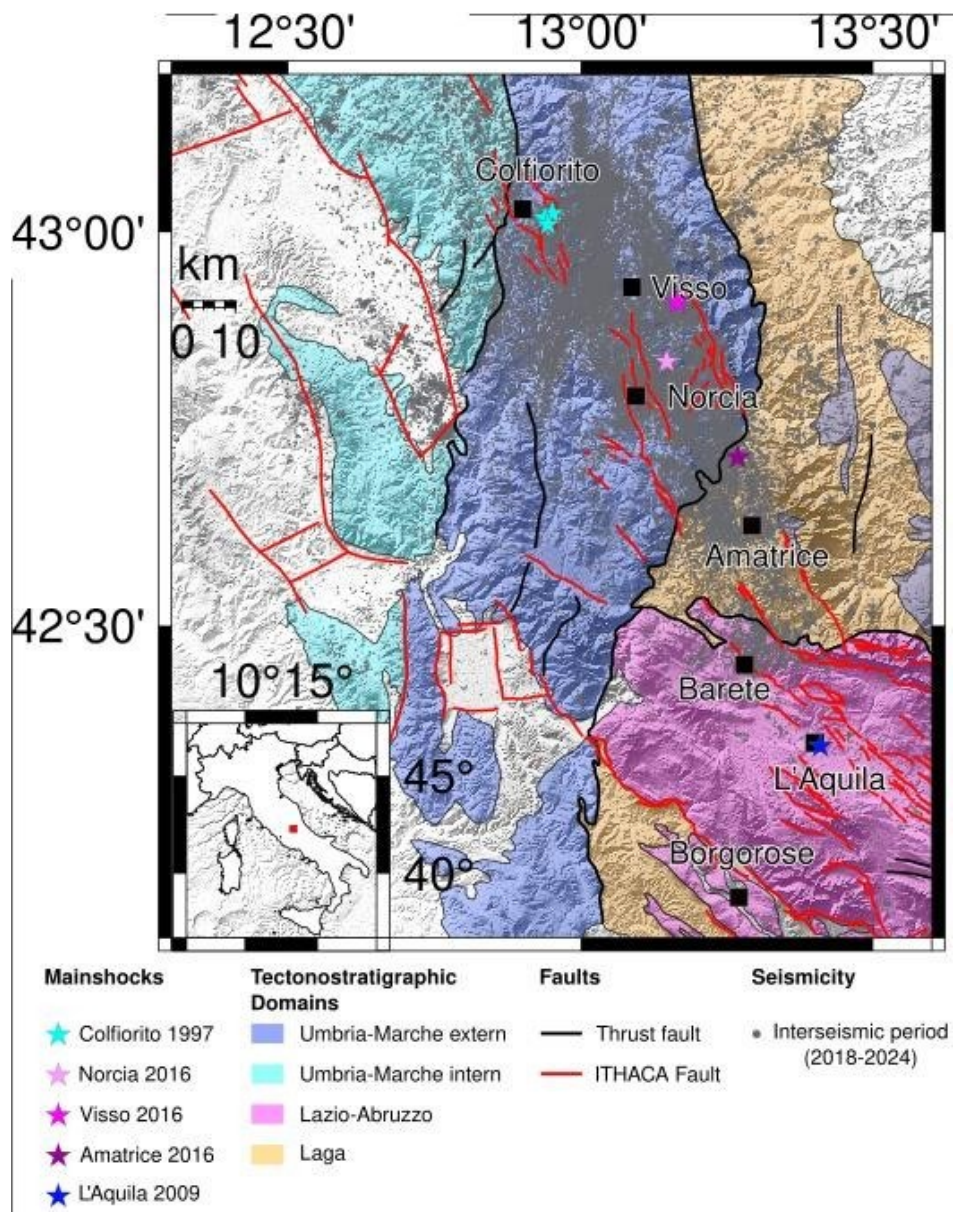

*Figure 1: Seismotectonic map of the Northern and Central Apennines (modified after Cosentino et al. (2010) and Porreca et al. (2018)). The grey dots show the relocated INGV catalogue. The red lines are the trace of active normal faults from ITHACA Working Group (2019). The black lines are the trace of the main thrust faults of the region. The stars indicate the epicentre of all earthquakes with magnitude greater than 5.5 that occurred in the area since 1997.*

The study area encompasses part of the Northern and Central Apennines (NCA), two sectors of a fold and thrust mountain belt developed since Late Miocene due to the post-collisional compression of the Adria continental passive margin (Bally et al., 1986; Centamore et al., 1993). The deformed sedimentary sequences record the long persistence of two large Meso-Cenozoic palaeogeographic domains: (i) the pelagic succession of the Umbria-Marche Basin (blue area in Fig. 1) and (ii) the Lazio-Abruzzo Platform (purple area in Fig. 1) consisting of Upper Triassic-Miocene shallow-water carbonates (e.g.,Bally et al., 1986; Buttinelli et al., 2021; Centamore et al., 1991; Cosentino et al., 2010; Ghisetti and Vezzani, 1998; Koopman, 1983; Lavecchia et al., 1988; Parotto & Praturlon, 1975). The transition between the two domains is represented by a band of proximal to distal slope deposits associated with widespread dolomitization along the Gran Sasso Range lato sensu (Ispra, 2012; Lucca et al., 2019, 2025). Such diverse sedimentary sequences were progressively involved in the Apennines orogeny during the eastward migration of the compressional front producing an overall thickness of the thrust sheet stack exceeding 8-10 km (Bally et al., 1986; Patacca et al., 2008; Scisciani et al., 2014; Porreca et al., 2018). While a combination of both thin- and thick-skinned tectonics is currently considered the most realistic model for the structure of the Umbria-Marche chain in the Northern Apennines, far less data and constraints are available for the Central Apennines (e.g., Barchi et al., 1998; Barchi & Tavarnelli, 2022; Ghisetti et al., 1993; Maresca et al., 2023). The major compressional tectonic contact is the Olevano-Antrodoco-Sibillini Mts. Line (also known as the Ancona-Anzio Line; e.g., (Salvini and Vittori, 1982), a regional fault system overthrusting the Northern Apennines over the Central Apennines. In addition, both the Northern and Central Apennines overthrust the Laga Domain (orange area in Fig. 1), composed of Upper-Miocene foredeep deposits (flysch), through arc-shaped major thrust systems (Barchi, Chiaraluce & Collettini, 2020). Following the last

Pliocene compressional pulses, the region experienced NE-SW oriented active extension (e.g., Ghisetti & Vezzani, 1991; Scandone et al., 1990), accommodated by Quaternary extensional fault systems responsible for the destructive seismic sequences of the last forty years (e.g., Chiarabba et al., 2009; Chiaraluce et al., 2017). Modern geodetic measurements reveal that extension (3-4 mm/yr on average) is concentrated on single fault systems in the north while it is distributed over a wider area and multiple fault systems in the Central Apennines (Daout et al., 2023). Coupling this observation with regional-scale seismic tomography of the region, suggest that these two portions of the mountain belt might be characterized by a significant difference in the upper crustal structure (Buttinelli et al., 2018). This difference may impact the spatio-temporal distribution seismicity in the Northern and Central Apennines.

To understand the processes and mechanisms driving the seismicity in a region, most studies also focused on seismic sequences, and more specifically on the Coseismic and early pre- and post-seismic phases (e.g., Chamberlain et al., 2021; Twardzik et al., 2022). This is understandable as there is sufficient seismicity to be able to observe and interpret the processes leading to the occurrence of a seismic sequence. On the other hand, the long-term study of Interseismic phases may help to understand the inner process that takes place during the quiescent period in the region under study. Therefore, we focused on the Interseismic period following the AVN sequence. Several questions arise: Is there seismic activity following the AVN sequence? If so, where is it located? Has it reactivated the same portions during the former seismic sequence? And is the seismicity of the same style during the Coseismic and Interseismic periods? In this paper, we try to answer these questions by first relocating the seismicity associated with the Interseismic period (2018 - 2024) and applying a declustering algorithm to define what type of seismicity occurred. We then analysed the spatio-temporal behaviour of this seismicity. Secondly, we compared the spatio-temporal behaviour of two subsequent Interseismic periods in the NCA region defined with respect to the L'Aquila and AVN seismic sequences. Finally, we compared the spatio-temporal behaviour of seismicity during Interseismic and Coseismic periods between the Northern and Central Apennines.

## 2. 2018-2024 Interseismic phase

We focus on seismicity between 42.1°- 43.2° N and 12.4° - 13.6° E (Figs. 1 and 2a). We specifically selected events that occurred between 01/01/2018 and 31/08/2024, in order to exclude the influence of the AVN seismic sequence. The seismic catalogue was retrieved from the INGV website (https://terremoti.ingv.it/), yielding a total of 55,278 detected events, with magnitudes ranging from ML -0.1 to 4.6 (Fig. 2). We refer to this as the original catalogue. Fig. 2(b) shows seismic activity in early 2018, coinciding with the strongest event of the period (ML 4.6). With the exception of this peak, the seismicity rate appears to be relatively stable throughout the 2018-2024 period (Fig. 2b).

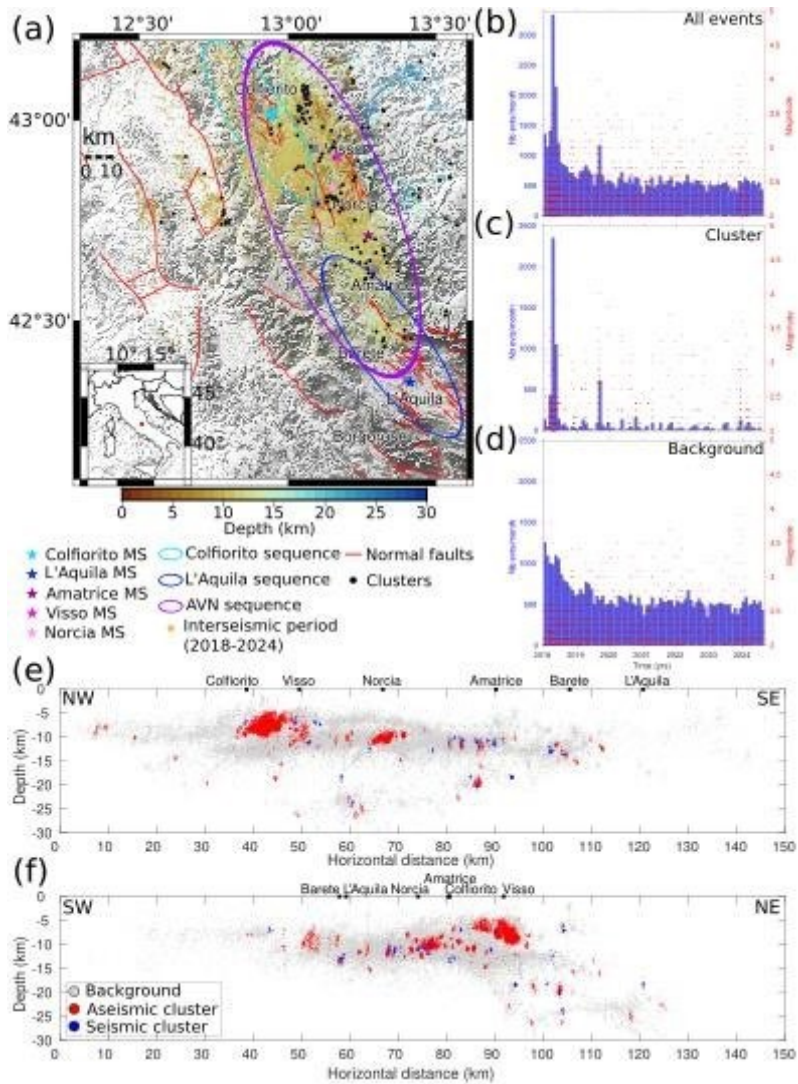

*Figure 2: Earthquake distribution for the 2018-2024 Interseismic period. In a), the coloured dots show the relocated INGV catalogue, with the colours from dark blue to dark orange showing the depth of the events. The light blue ellipse includes the earthquake location of the 1997 Colfiorito sequence, the blue ellipse the 2009 L'Aquila sequence, and the purple ellipse includes the 2016-2017 Amatrice-Visso-Norcia sequence. The coloured stars show the location of the strongest events in the Coseismic sequences. The red lines show the active normal faults from ITHACA Working Group (2019). The black dots show the average location of the clusters in the 2018-2024 Interseismic period. In b), the temporal distribution of the relocated INGV catalogue. In c), the temporal distribution of the events occurring in clusters. In d), the temporal distribution of the events occurring in the background. The red dots show the*

*magnitude of events greater than 2 in b), c), and d). In e) and f), the projection of all the Interseismic period on vertical planes (cross-section). In grey, the background events; in red, the aseismic cluster events; in blue, the seismic cluster events.*

### 2.1. Data

To analyse the spatial distribution of the seismicity, it was first relocated using data from 196 stations operating within 200 km distance of events, throughout the 2018-2024 period (Fig. S1). The hypoDD software (Waldhauser & Ellsworth, 2000), based on the double-difference technique, was used to refine the event locations. The time catalogue selected by INGV staff for each event was used as input.

Several 1D velocity models have been proposed for this region ((Bagh et al., 2007; Chiarabba et al., 2009; Cirella et al., 2009; Chiaraluce et al., 2011; Herrmann, Malagnini & Munafò, 2011; Carannante et al., 2013). To assess their impact on earthquake relocation, we re-located the original catalogue using these different velocity models, with the relocation result from Chiaraluce et al. (2011) serving as our reference. This model was also used by Valoroso et al. (2013) in the relocation of the 2009 L'Aquila sequence. For each velocity model, we calculated the differences in position (longitude, latitude, and depth) relative to the relocation based on the Chiaraluce et al. (2011) model (see Table S1). The differences were minimal, typically less than 150 meters, for all components (longitude, latitude, and depth). This indicates that the different velocity models yield very similar relocations in this region. To ensure consistency with previous studies, we therefore chose the Chiaraluce et al. (2011) velocity model for this work.

Approximately 98% of the events were successfully relocated, amounting to 54,151 earthquakes. On average, seismicity was found to be shifted slightly (~0.06 km westward, ~0.04 km southward, and ~0.25 km shallower) compared to the original catalogue (Fig. S2). Despite these minor differences, the overall relocated positions are consistent with the original data (Fig. S2).

## 2.2. Declustering

On the newly relocated data, we apply the declustering algorithm of Zaliapin & Ben-Zion (2020) that uses the nearest-neighbour approach in space-time-magnitude domain. We first plotted the joint distribution of the rescaled time (T) and distance (R) components of the nearest-neighbour distance η (eq.1)

$$log\eta_{ij} = logT_{ij} + R_{ij} (1)$$

With event i, the parent event of event j.

As we observed a bimodal distribution on Fig. S3(a), we can apply the declustering algorithm. We define the fractal dimension to be 1.9, the threshold separating background and cluster event $\eta_0$ of -3.4 (limit of the bimodal distribution, see Fig. S3b), and b=0 as defined by Zaliapin & Ben-Zion (2020).

In total, 38 192 clusters were found, with 1 to 1373 events. We only consider a cluster to be one if it has at least 10 events. Of the 38 192 clusters, 187 have between 10 to 1373 events (black dots, Fig. 2a). They last from ~3 hours to ~291 days and are distributed over areas from ~0.02 to ~93 km², between 5 and 27 km depth. Of the 54 151 relocated events, 7 628 events (~14%) are included in the clusters, with magnitude 0<=ML<=4.6 (Fig. 2c).

## 2.3. Global analyses

The temporal distribution of the events belonging to the cluster is shown in Fig. 2(c). There is a strong cluster activity at the beginning of 2018, which may be related to the ML 4.6 earthquake registered in April of that year. There is a second strong cluster activity at the end of 2019, also associated with an ML> 4 event. Apart from these two strong activities, the cluster activity appears to be generally low (< 200 events/month), with only some small peaks (Fig. 2c). Compared to the background activity (Fig. 2d), the cluster activity slowly decreased from 01/01/2018 until it reached a stable level of ~550 events/months in the mid-2019.

In Figs. 2(e) and 2(f), we have projected the seismicity onto vertical planes (cross-sections) oriented NW-SE and SW-NE, respectively. Looking at the cross sections (Figs. 2e and f), it appears that the volume of the clusters is smaller towards the southeast (Fig. 2e). A similar observation is made on the SW-NE cross-section (Fig. 2f). We also observe a low-angle structure on both cross-sections, localized between 7 and 15 km depth, which is mainly highlighted by the background seismicity (Figs. 2e and f). Most of the seismicity (93%) occurred on or above this low-angle structure. Cluster seismicity also mostly occurred above 15 km, with 93% of the events cluster and corresponding at 88% of the clusters. Interestingly, this cloud of low-angle seismicity disappears in the south-eastern part of the region, which is affected by the presence of active seismogenic faults (i.e. the SE portion of the cross-section in Fig. 1(e) running parallel to the average strike of the active faults).

## 2.4. Behaviour of seismicity

To gain insights about the physical processes driving the seismicity in each cluster, we used a set of parameters bearing information about their spatio-temporal evolution and magnitude distribution. The first parameter is the seismic-to-total moment ratio (Danré et al., 2022), which provides insight into the relative contribution of seismic versus aseismic processes. When this ratio is close to 1, it indicates that the seismic process is dominant in controlling the observed seismicity. This ratio is expressed as:

$$Ratio = \frac{Seismic\, cumulated\, moment}{Total\, moment} \quad (2)$$

$$Total\, moment = G * Dmax * A \quad (3)$$

With the shear-modulus G (Pa); the average slip over the event area Dmax (m); the area of the cluster, A (m²). The area of the cluster is obtained by selecting the events that define the periphery of the cluster and calculating the internal area. Is worth mentioning that this is an apparent area in case of volumetric distribution of seismicity or if earthquakes are located in near vertical faults, and we explore and discuss this effect in the rest of the manuscript.

To estimate Dmax, we assume that the largest event has a circular rupture with an area A, shear modulus G of 30 GPa (a typical value for crustal rocks), and a static stress-drop of 10 MPa (Madariaga, 1976). The slip is given by (Madariaga, 1976):

$$Dmax = \frac{(16\Delta\sigma max)^{2/3} * (Seismic\, moment\, max)^{1/3}}{(7)^{2/3} G \pi} \quad (4)$$

With Δσmax, the maximum static stress-drop (Pa).

Another important parameter is the effective stress-drop (Fischer & Hainzl, 2017), which can provide insight into the role of aseismic processes in controlling seismicity. A low effective stress-drop (< 0.1 MPa) typically indicates that the seismicity is sparsely distributed, suggesting the presence of aseismic slip contributing or driving the seismicity. In contrast, a high effective stress-drop implies that the "seismic" asperities cover a larger portion of the seismic area, which is in turn associated with a lower contribution from aseismic processes.

The effective stress-drop is defined as:

$$\Delta\sigma e = \frac{7 * Seismic\,cumulated\,moment}{16 * Ra^3} \quad (5)$$

With Ra (m), the radius of the area A of the cluster.

The radius Ra may be expressed as:

$$Ra = \sqrt{\frac{A}{\pi}} \quad (6)$$

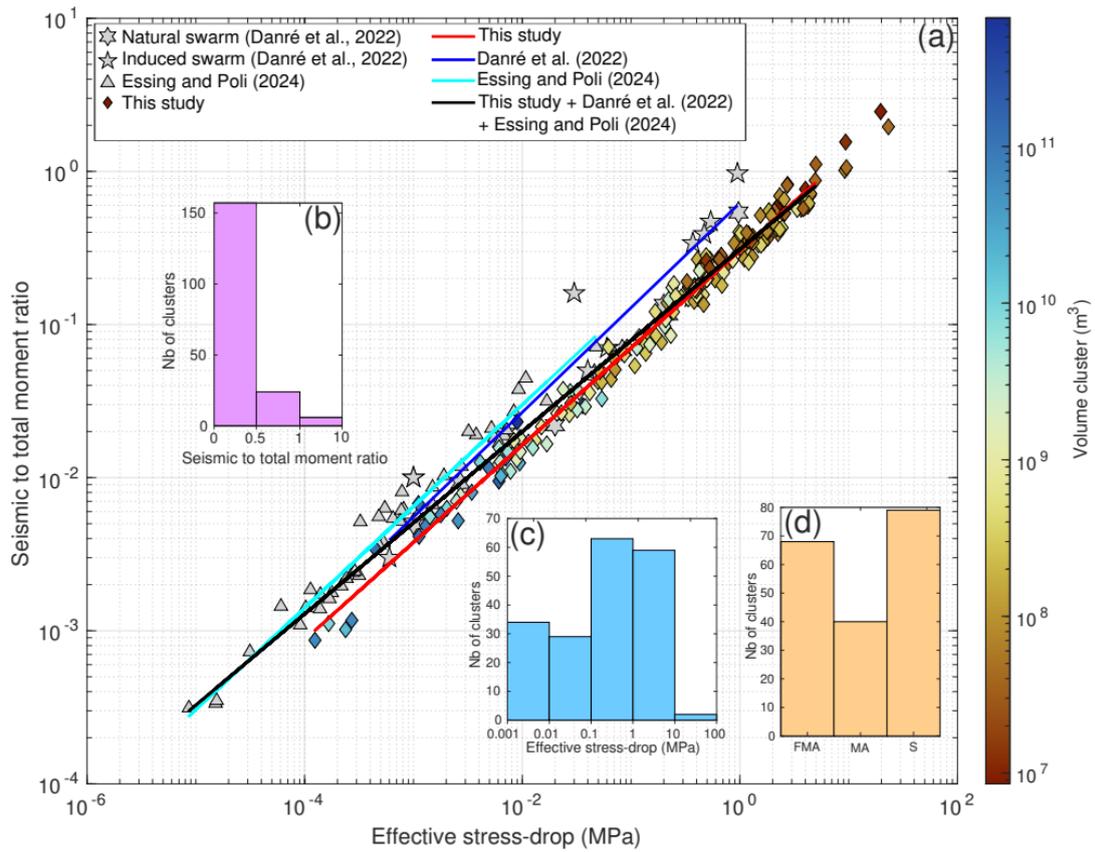

*Figure 3: Seismic to total moment ratio as function of the effective stress-drop. In a), the diamonds correspond to the cluster of this paper, the triangle to Essing & Poli (2024) study, and the stars to Danré et al. (2022). The several lines correspond to the best-linear fit for different datasets. For the best-linear fit of datasets including the clusters of this paper, we do not include the clusters that have a ratio > 1. The colour of the diamond corresponds to the volume occupied by the events in the cluster. It was calculated using the boundary events of the cluster. In b), histogram of the Seismic to total moment ratio. In c), histogram of the effective stress-drop in MPa. In d), histogram of the Ogata & Katsura (2012) classification for the 2018-2024 Interseismic period cluster.*

We plot the ratio as a function of the effective stress-drop for the 187 clusters in Fig. 3(a). Of the 187 clusters, 157 (84%) have a ratio lower than 0.5 and 24 (13%) have a ratio between 0.5 and 1 (Fig. 3b). Of the 157 clusters with a low ratio, 126 (67%) have an effective stress-drop lower than 1 MPa. Of the 24 clusters with a high ratio, all of them have an effective stress-drop greater than 1MPa. The relationship between the effective stress-drop and the ratio can be described by a linear fit. On the other hand, there are 6 clusters with a ratio > 1, with less than 14 events. All of them also have an effective stress-drop greater than 4MPa. The high ratio and the effective stress-drop might indicate a mainshock-aftershock type of activity.

Ogata & Katsura (2012) proposed a classification that defines whether a cluster is associated with mainshock-aftershocks (MA), foreshocks-mainshocks-aftershocks (FMA) or swarms (S) activity. In MA sequences, the strongest event in the cluster occurs first in time. To distinguish FMA sequences from S sequences, the difference (DM) between the magnitude of the strongest pre-shock event (occurring before the mainshock) and the mainshock must be greater than 0.45. If so, the cluster is qualified as associated with an FMA sequence. If DM < 0.45, then the cluster is associated with a swarm (S) activity. Following this scheme, in Fig. 3(d), we classify the clusters as 42% swarms (S), 36% foreshock-mainshock-aftershock (FMA), and 21% mainshock-aftershocks (MA).

In the following, the clusters are divided into two categories (aseismic and seismic) based on the ratio and effective stress-drop. A cluster was classified as aseismic if it had a ratio < 0.5 and an effective stress-drop < 1MPa, while a cluster was classified as seismic if it had a 0.5 <= ratio <= 1.

## 2.5. Spatial behaviour of seismicity

Looking at the location of the clusters on the map (Fig. 4), we select six areas where most of the clusters occurred. Moving from NW to SE, the first area is near Colfiorito (light blue rectangle, Fig. 4), the second one around Visso (blue rectangle, Fig. 4), the third around Norcia (dark blue rectangle, Fig. 4), the fourth around Amatrice (purple rectangle, Fig. 4), the fifth one near the village of Barete (magenta rectangle, Fig. 4), and the last one close to town of L'Aquila (pink rectangle, Fig. 4). Each line in Fig. 4 represents the cross-sections in Fig. 5. The dashed-lines on Fig. 4, show these areas projected onto different ~SW-NE oriented cross-sections (Fig. 5).

It was observed that there were fewer clusters in the Barete and Amatrice areas than in Norcia, Visso and Colfiorito areas (Fig. 5). The clusters were also more seismic (see definition above) in the southern areas (Barete and Amatrice) than in the northern areas (Norcia, Visso, Colfiorito). The clusters became deeper as they moved from northwest to southeast (Fig. 5). In particular, in the Colfiorito area (Fig. 5a), the clusters occurred mainly above the low-angle seismicity cloud previously defined (Figs. 5b, c, d, and e), while in the other areas, the clusters are located within such low-angle seismicity volume. Importantly, as a matter of fact, no clusters (see Figs. 2, 4, and 5) were detected in the L'Aquila area (the southernmost portion of the studied region).

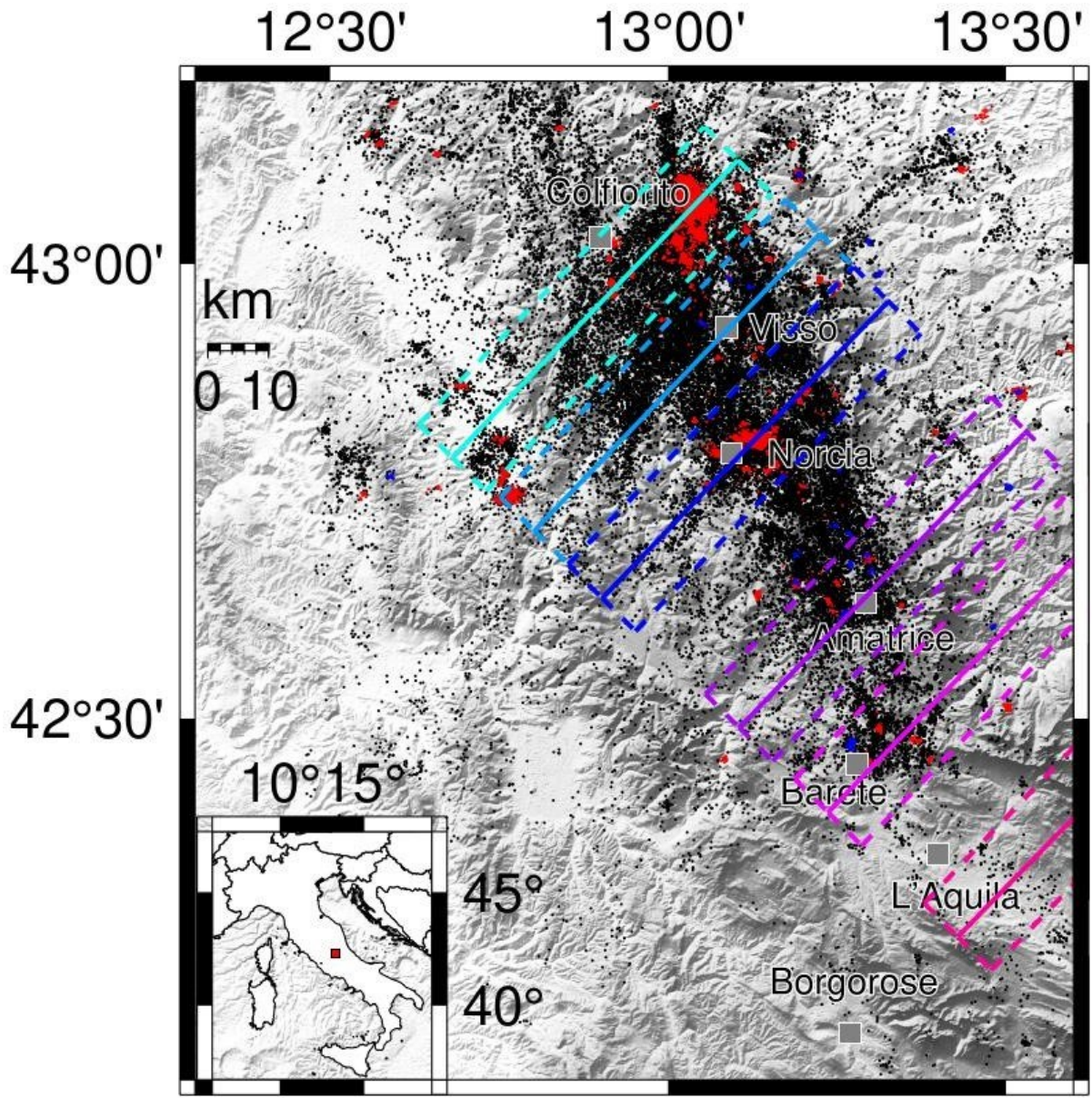

*Figure 4: Location of cross-sections. The black dot corresponds to the background events; the red dot to the aseismic cluster events; the blue dot to the seismic cluster events. The lines correspond to the traces of cross-section in Fig. 5. The dashed-line shows the area of projection in the cross-section in Fig. 5.*

303

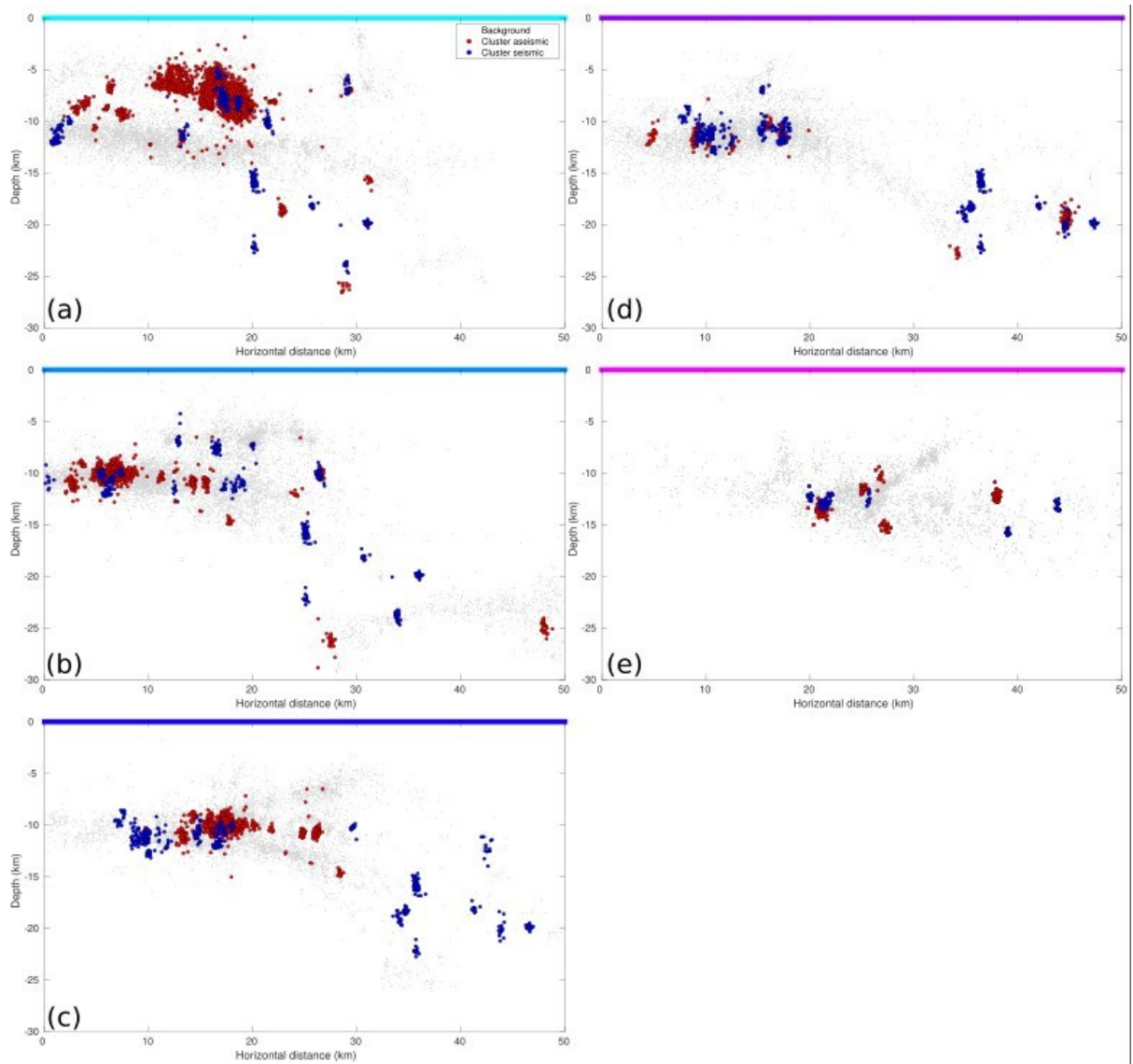

*Figure 5: Cross-section for 5 specific areas. The black dot corresponds to the background events; the red dot to the aseismic cluster events; the blue dot to the seismic cluster events. The coloured lines correspond to the traces of cross-section in Fig. 4. In a), area around Colfiorito; in b), area around Visso; in c), area around Norcia; in d), area around Amatrice; in e), area around Barete; in f), area around L'Aquila*

## 3. Interseismic periods (2011-2016 and 2018-2024)

So far, we have focused our observations on the Interseismic period following the AVN sequence. With our observations, we can now compare our results with the previous Interseismic period that occurred in the region between the L'Aquila and AVN sequences (2011-2016). We refer to the post-AVN Interseismic period as the 2018-2024 Interseismic period and to the previous between L'Aquila and AVN sequences as the 2011-2016 Interseismic period.

### 3.1. 2011-2016 Interseismic period

Sugan et al. (2023) relocated the ~23,000 events from the 1rst January 2009 to the 24th of August 2016. They used template-matching to increase the number of events and created a catalogue of ~114,000 events. From their catalogue, we keep only those events that occurred after the 1rst January 2011, to remove the effect of the 2009 L'Aquila sequence. We call this catalogue, the 2011-2016 Interseismic period. Most of the seismicity is located below the 7 km depth (Figs. 6b, 7b, and S4). Fig. S4 shows the catalogue of Sugan et al. (2023) in light blue and our relocated catalogue in dark blue. Of the 114,000 events, 51% are within a cluster while only 17% of the seismicity for the 2018-2024 Interseismic period is within a cluster.

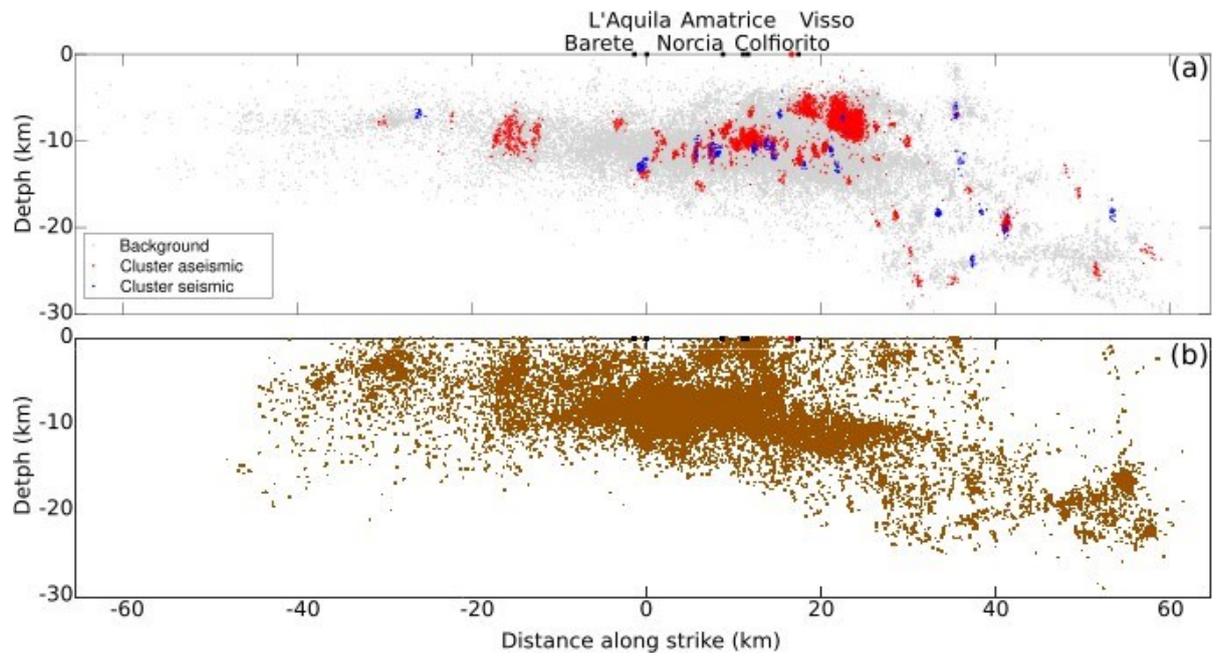

*Figure 6: Cross-section oriented N55° (perpendicular to active extensional fault system). In a) the 2018-2024 Interseismic period. In b), the 2011-2016 Interseismic period (Sugan et al., 2023).*

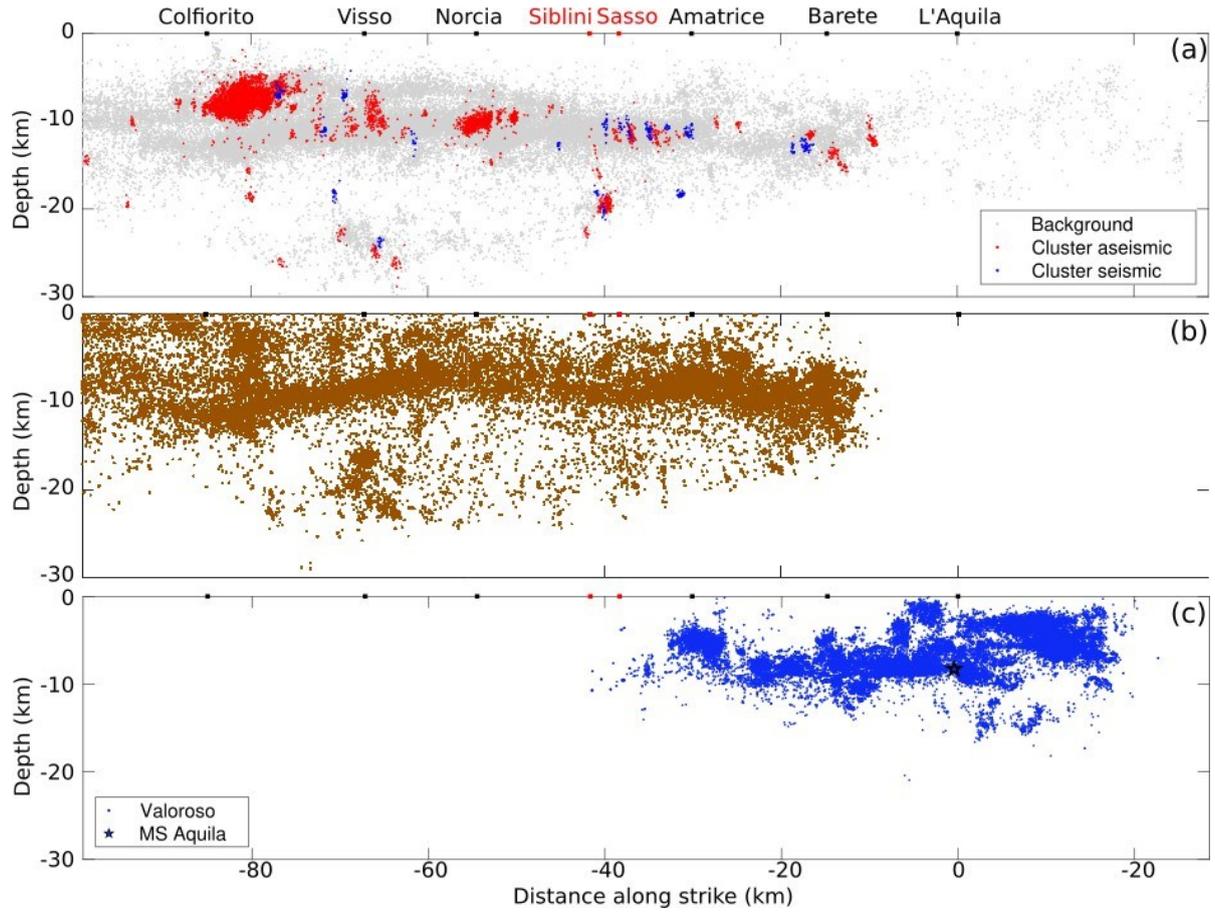

Figure 7: *Cross-section oriented N145° (parallel to the active fault system). In (a) the 2018-2024 Interseismic period. In b), the 2011-2016 Interseismic period (Sugan et al., 2023). In c), the L'Aquila sequence from Valoroso et al. (2013).*

## 3.2. Seismicity location

In Fig. 6, we present the projection of the seismicity on a vertical plane (cross-section) oriented N55° for several catalogues: in a, the 2018-2024 Interseismic period; in b, the 2011-2016 Interseismic period. Fig. 7 also shows the projection of the seismicity on a vertical plane (cross-section) oriented N145° for several catalogues: in a, the 2018-2024 Interseismic period; in b, the 2011-2016 Interseismic period; and in c, the 2009 L'Aquila sequence (Coseismic period). It is very interesting to note that the same type of structures is highlighted by the seismicity for both Interseismic periods (Figs. 6a b, 7a, and b). The only difference is the depth at which the seismicity occurs. A possible explanation for the depth difference might be that we have not taken into account the topography in our relocation. If we subtract the average topography (1000 m) from the depth of our events, we better agree with the depth of Sugan et al. (2023). Furthermore, the relocation obtained by hypoDD is the relative position between events, which means that there is an uncertainty in the absolute position. Then, the combination of topography and location uncertainty might then explain the depth difference.

### 3.3. Cluster comparison

Sugan et al. (2023) also applied the declustering algorithm of Zaliapin & Ben-Zion (2020) to their catalogue. They found 670 clusters with at least 10 events, lasting from days to months with maximum local magnitude < 3. Some of their clusters are localized between the town of Amatrice and the Amatrice mainshock. We also find clusters in this area for the 2018-2024 Interseismic period (Fig. 5d). However, compared to the 2018-2024 period, we find fewer clusters occurring to the north, close to Visso (Figs. 5a and b). It seems that the location of the general seismicity is very similar, but when looking at the clustered seismicity, the locations may change.

### 3.4. Type of seismicity

Sugan et al. (2023) also used the Ogata & Katsura (2012) classification and found that the clusters were classified as 55% as swarm (S), 43% foreshock-mainshock-aftershock (FMA) and 1% mainshock-aftershock (MA). The classification for the 2018-2024 period showed similar FMA (36%) and S (42%), however, there was a larger proportion of MA (21%) than for the 2011-2026 Interseismic period.

## 4. Coseismic and 2018-2024 Interseismic periods

The NCA region has been affected by 3 strong and destructive seismic sequences: 1997 Colfiorito, 2009 L'Aquila and 2016-2017 Amatrice-Visso-Norcia. Each of these sequences will be referred to here as a "Coseismic period". Below, we will describe each Coseismic period and compare their behaviour with the respective Interseismic periods.

### 4.1. 1997 Colfiorito sequence

The 1997 Colfiorito sequence (light blue ellipse, Fig. 2a) began on 3 September 1997 with a Mw 4.5 and was followed by a low seismic activity (Ripepe, Piccinini & Chiaraluce, 2000). Three weeks later, on 26 September 1997, a Mw 5.7 event occurred, followed nine hours later by a stronger event (Mw 6.0). Then, on 14 October, an Mw 5.6 event was generated (corresponding to the last strong event in the sequence). Between 26 September and 3 November 1997, more than 2,000 events were detected (Deschamps et al., 2000). Deschamps et al. (2000) that the seismic moment released by the events of 26 September represented slightly more than 50% of the total moment released during the sequence. According to Scholz (1990), this is less than would be expected from a mainshock-aftershock sequence. Furthermore, Deschamps et al. (2000) observed three areas with a different seismic behaviour: mainshock-aftershocks, cluster seismicity, and large events (Mw > 4) occurring during almost all the sequence. The seismicity was mainly located above 9 km depth (Deschamps et al., 2000; Chiaraluce et al., 2005).

## 4.2.  2009 L'Aquila sequence

The L'Aquila sequence (blue ellipse, Fig. 2a), which started in January 2009, generated more than 5,000 events before the Mw 6.1 mainshock on 6 April 2009, followed by more than 50,000 aftershocks (Valoroso et al., 2013). Valoroso et al. (2013) used automated P- and S-waves detection coupled with cross-correlations to relocate the seismicity using a double-difference algorithm. More than 64,000 events were relocated with a completeness magnitude of 0.7. Most of the seismicity was located above 12km, with one group of events associated with a fault occurring at ~ 15 km (Fig. 7c). 425 clusters with a cross-correlation coefficient > 0.96. These clusters contained between 3 and 24 events.

## 4.3.  2016-2017 Amatrice-Visso-Norcia sequence

Seven years after the L'Aquila sequence, the AVN sequence occurred (purple ellipse, Fig. 2a), starting with a Mw 6.0 event on 24 August 2016 and culminating in a Mw 6.5 earthquake on 30 October 2016, with over 450,000 events recorded in the year following the mainshock (Spallarossa et al., 2021).

Tan et al. (2021) used a deep-neural-network-based phase picker (Zhu & Beroza, 2019) and a double-difference relocation algorithm (Waldhauser & Ellsworth, 2000) to relocate over 900,000 events (magnitude of completeness > 0.3) for a one year starting on 15 August 2016. The seismicity is located above 12 km.

Volpe et al. (2023) applied a declustering method to the relocated catalogue of Tan et al. (2021) and used the ST-DBSCAN algorithm (Birant & Kut, 2007) to perform spatio-temporal clustering. In the case of Volpe et al. (2023), the events were only assigned to a cluster if they had a maximum-neighbour-distance of less than 150 m and occurred less than 14 days after the first event in the cluster. They obtained 625 clusters, representing 15% of the seismicity, and occurring between 9 to 12 km depth.

### 4.4. Coseismic and Interseismic periods comparison

Most of the Coseismic events are located above 13 km depth, although the Colfiorito events are located shallower than the L'Aquila and AVN events. The areas activated by the Coseismic sequences of Colfiorito and AVN are still active during both Interseismic periods. On the other hand, the area of the L'Aquila sequence was apparently "silent" during both Interseismic periods.

During the Coseismic periods, the sequences are mainly of mainshock-aftershocks style with some foreshock-mainshock-aftershock. However, some studies have shown that the Coseismic sequences also have clustered seismicity. It seems that the seismic behaviour of the Coseismic and Interseismic periods is similar. Indeed, for the 2011-2016 and the 2018-2024 Interseismic periods, foreshock-mainshock-aftershocks type of sequences are found for 43% (2011-2016) and 36% (2018-2024) of the clusters. This type of sequence is also observed for the Colfiorito sequence (Sebastiani, Govoni & Pizzino, 2019) and L'Aquila sequence (Cabrera, Poli & Frank, 2022). Swarm-like seismicity is also found for the Interseismic and Coseismic periods. 55% (2011-2016) and 42% (2018-2024) of clusters in the Interseismic periods are classified as Swarm and some areas during the Colfiorito sequence showed clustered seismicity. However, even if some clusters in the Interseismic periods (1% for 2011-2016 and 21% for 2018-2024) show mainshock-aftershocks behaviour, this is not the main type of seismicity compared to the Coseismic sequences, which are mainly driven by mainshock-aftershock sequences.

## 5. Discussion

### 5.1. Processes

The seismic-to-total moment ratio and effective stress-drop obtained for each cluster during the 2018-2024 Interseismic period may be compared with other seismic cluster studies. Indeed, Danré et al. (2022) calculated the seismic-to-total moment ratio and effective stress-drop for several natural and induced swarms (Fig. 3a, stars). We also calculated these parameters for the 2012-2017 Alto-Tiberina cluster catalogue from Essing & Poli (2024), located NW of the Colfiorito area. Danré et al. (2022) showed that there is a log-linear relationship between the seismic-to-total moment ratio and the effective stress-drop. The clusters for the 2018-2024 Interseismic period have similar values of seismic-to-total moment ratio and effective stress-drop as Danré et al. (2022) and Essing & Poli (2024) (Fig. 3a). Since more than half of the clusters (67%) have a seismic-to-total moment ratio lower than 0.5 and an effective stress-drop lower than 1MPa, the main process driving the seismicity during the 2018-2024 Interseismic period is aseismic.

## 5.2. Type of seismicity

Several papers have defined clusters for the 3 Coseismic periods. Deschamps et al. (2000) observed an area of clustered seismicity for the 1997 Colfiorito sequence. Valoroso et al. (2013) also observed clustered seismicity defined by a high cross-correlation coefficient for the 2009 L'Aquila sequence. Volpe et al. (2023) also defined seismic clusters for the 2016-2017 AVN sequence. All these different clusters found for the Coseismic periods are difficult to compare as they have been characterised differently. Furthermore, no study on the temporal, spatio-temporal or the Ogata & Katsura (2012) classification study was performed on these clusters, making it difficult to define whether these clusters behave like a swarm or more like a foreshock-mainshock-aftershocks. Nevertheless, clusters are found for each Coseismic period. As clusters have been defined for both Interseismic periods, this means that there is clustered seismicity in the NCA region regardless of the period.

Regarding the type of seismicity occurring in the NCA region, Cabrera et al. (2022) studied the foreshocks that preceded the L'Aquila mainshock, proving that there is another type of seismicity other than mainshock-aftershocks for the 2009 L'Aquila sequence. For the Interseismic periods, following the classification of Ogata & Katsura (2012), Sugan et al. (2023) and ourselves showed that the Interseismic periods are mainly struck by foreshocks-mainshocks-aftershocks (43% for 2011-2016 and 36% for 2018-2024) and swarms (55% for 2011-2016 and 42% for 2018-2024), with fewer mainshock-aftershock sequences (1% for 2011-2016 and 22% for 2018-2024). The difference between Coseismic and Interseismic period seems to lie in the dominant type of seismicity: in the Coseismic period, it is mainly mainshock-aftershocks, while in the Interseismic period it is foreshocks-mainshocks-aftershocks and swarms. This difference is highlighted by the maximum magnitude of the mainshocks. For the Interseismic periods, the maximum magnitude for the 2011-2016 Interseismic periods is ML 4 and ML 4.6 for the 2018-2024 Interseismic period, while for the Coseismic periods, it is higher than ML5.5.

### 5.3. Same sites during Interseismic periods

Comparing the location of the seismicity for both Interseismic periods, we found that the seismicity seems to illuminate the same structures (Figs. 6a,b, 7a, b, and S4). As briefly discussed in the Interseismic periods (2011-2016 and 2018-2024) section, there is a difference in the depth of the location of the seismicity. This difference may be due to the location uncertainty and to the average topography of the NCA region. However, both Interseismic periods activate the same areas. This might mean that the processes driving the Interseismic period are similar each time. On the other hand, when comparing the location of the clusters, it seems that they are not located in the same place between the two Interseismic periods. This means that there is something specific to each Interseismic period that drives the seismicity but does not affect the general seismicity. One process might be the fluids at depth that migrate through the medium in the NCA region as several studies have inferred their presence at depth in different locations (Terakawa, Miller & Deichmann, 2012; Poli et al., 2020; Chiaraluce et al., 2022).

### 5.4. Cluster areas

From the cross-sections in Figs. 2(e) and (f), it appears that the volume of clusters decreases towards the southeast. In parallel, the number of clusters also decreases towards the southeast, as can be seen from the cross-sections in Fig. 5, where there are fewer clusters in Amatrice (Fig. 5d) and Barete (Fig. 5e) than in Colfiorito, Visso and Norcia (Figs. 4a, b, and c, respectively). The difference in the number of clusters might be due to the different structure and composition of the upper crust in the Amatrice and Barete areas compared to northernmost regions. At least, in the case of the Amatrice area, the upper 2-3 km are characterized by the presence of foredeep deposits of the Laga Unit (Fig. 1; Porreca et al., 2018).

The clusters located in the areas of Amatrice, Barete, Norcia, and Visso (Figs. 5b, c, d, and e) areas seem to occur inside the low-angle seismicity cloud defined by the background events rather than above, as it is the case for the Colfiorito area (Fig. 5a).

### 5.5 Difference North-South

Another important observation is the absence of seismicity in the L'Aquila area during both Interseismic periods and during the AVN sequence. The quiescence observed for the 2011-2016 Interseismic period may be explained by the fact that the area is completely stress-free following the 2009 L'Aquila sequence. The inactivity observed for the AVN sequence may also be explained by the absence of stress still following the L'Aquila sequence. However, it is difficult to explain why the area is still quiet after more than 10 years, whereas the Norcia and Colfiorito areas were active during both Interseismic periods. This is the ones of the difference between the north and south of our study area.

As a matter of fact, we showed that there was less seismic activity in the south part (L'Aquila - Barete - Amatrice area) than in the north (Colfiorito – Visso – Norcia area) (Fig. 5). The lack of seismic activity in the south was also observed for the 2011-2016 Interseismic and AVN Coseismic sequences. Moreover, we also found that there was a difference in the type of clusters. There were more clusters aseismic in the north than in the south, and more clusters seismic in the south than in the north (Fig. 5). Additionally, we also found that the volume of the cluster decreased towards the south (Figs. 2e and 5). Furthermore, we also noticed that the clusters occurred at deeper depth in the south than in the north (Fig. 5). It is also observed for the Coseismic sequences, where the Colfiorito events occurred at shallower depth than the AVN and L'Aquila sequences. These differences observed between the north and the south by seismology are also observed by other studies. As we stated in the Introduction section, the structure of the upper crust in the study region differs significantly between Northern and Central Apennines, being two distinct palaeogeographic domains since the Meso-Cenozoic later involved in the mountain belt build up. In particular, some authors provide evidence of a dolomitized ultrathick carbonate sequence dominating the crust of the L'Aquila region (Minelli & Speranza, 2015; Buttinelli et al., 2018) in contrast with the Umbria-Marche domain to the north. For instance, based on regional scale seismic tomography (Buttinelli et al., 2018) and (Fonzetti et al., 2025) showed the occurrence of high-velocity (Vp up to 7 kms$^{-1}$) bodies between 4 and 12 km depth contrasting with the significantly lower velocities retrieved at the same depths to the north in the Alto-tiberina, Colfiorito and Norcia-Visso region. Such contrasting crustal structures might play a pivotal control on the long-term active deformation and seismicity in the area.

## 6. Conclusion

In this paper, we have chosen to study the 2018-2024 Interseismic period in the Northern-Central Apennines. To do this, we first relocated the seismicity and applied a declustering algorithm to define clusters. We found that most of the clusters are mainly driven by an aseismic process and that the volume of the clusters decreases towards the southeast of the NCA region while the depth of the clusters increases. We also compared our Interseismic period with the 2011-2016 Interseismic period and found that the Interseismic events appear to be located at the same locations. The clusters in the Interseismic periods are mainly characterized by foreshocks-mainshocks-aftershocks and swarm sequences compared to mainshock-aftershocks sequence that leads the Coseismic periods. We also observed a difference in the seismicity behaviour between the north and south of our study area, with more aseismic clusters, bigger clusters, shallower seismicity in the north compared to the south. That difference may be due to the composition of the rocks.

Extending the 2018-2024 Interseismic catalogue by template-matching and starting at the beginning of the AVN sequence might help us to better understand the seismic behaviour during both Coseismic and Interseismic period in the NCA area.

## Acknowledgments

Most of this work was done thanks to the type A Research grant allocating to M. Baques from the University of Padova.## Author Contributions:

Conceptualization: Marion Baques, Piero Poli, Michele Fondriest

Formal analysis: Marion Baques, Piero Poli, Michele Fondriest

Funding acquisition: Piero Poli, Michele Fondriest

Investigation: Marion Baques, Piero Poli, Michele Fondriest

Methodology: Marion Baques, Piero Poli

Project Administration: Piero Poli, Michele Fondriest

Supervision: Piero Poli, Michele Fondriest

Validation: Marion Baques, Piero Poli, Michele Fondriest

Visualization: Marion Baques

Writing – original draft: Marion Baques

Writing – review & editing: Marion Baques, Piero Poli, Michele Fondriest

588

## Data Availability

Software: The figures were created using GMT (Wessel et al., 2019), Matlab (MathWorks, 2023), and the Scientific ColorMaps from Crameri (2018). The declustering algorithm comes from [https://github.com/dttrugman/Nearest_Neighbor_Declustering/tree/main](https://github.com/dttrugman/Nearest_Neighbor_Declustering/tree/main).

Data: The INGV catalogue is available on the website (https://terremoti.ingv.it/). The relocated catalogue and cluster catalogue are available online. All the seismic data catalogues used in this paper are available for each article from which they originate.

## Supporting Information

Supplementary figures and data:

Supplementary_materials.docx

Catalogue_events_20180101_20240831.zip

Catalogue_information_clusters.zip

## Conflict of Interests

The authors declare no conflicts of interest relevant to this study.